\documentclass{iopart}
\usepackage{iopams}
\usepackage{dsfont}
\usepackage{amsthm}
\usepackage{graphicx}
\usepackage{cite}

\newcommand{\uP}{\,\mathbb{P}}
\newcommand{\ud}{\,\mathrm{d}}
\newcommand{\ue}{\mathrm{e}}
\newcommand{\ui}{\mathrm{i}}
\newcommand{\uj}{\mathrm{j}}
\newcommand{\uk}{\mathrm{k}}
\newcommand{\RR}{\mathds{R}}
\newcommand{\HH}{\mathds{H}}

\def\Re{{\rm Re}}
\def\Im{{\rm Im}}

\begin{document}
\title{Random matrix ensembles for $PT$-symmetric systems}
\author{Eva-Maria Graefe and Steve Mudute-Ndumbe and Matthew Taylor}
\address{Department of Mathematics, Imperial College London, London, SW7 2AZ, United Kingdom}
\begin{abstract}
Recently much effort has been made towards the introduction of non-Hermitian random matrix models respecting $PT$-symmetry. Here we show that there is a one-to-one correspondence between complex $PT$-symmetric matrices and split-complex and split-quaternionic versions of Hermitian matrices. We introduce two new random matrix ensembles of (a) Gaussian split-complex Hermitian, and (b) Gaussian split-quaternionic Hermitian matrices, of arbitrary sizes.  We conjecture that these ensembles represent universality classes for $PT$-symmetric matrices. For the case of $2\times2$ matrices we derive analytic expressions for the joint probability distributions of the eigenvalues, the one-level densities and the level spacings in the case of real eigenvalues. 
\end{abstract}

In recent years there has been a surge of research interest in $PT$-symmetric quantum theories, accompanied by a multitude of experimental applications and realisations \cite{Bend98, 10ruet,Schi11,12rege,Bitt12,13feng, 14bran, 14peng, 14sun, 14chan,14feng, 14hoda, 14peng2, 12brod,Scho10,ElGa07,Guo09}. For finite-dimensional systems represented by matrices, $PT$-symmetry is equivalent to the reality of the characteristic polynomial \cite{10bend}. That is, $PT$-symmetric matrices have either real or complex conjugate eigenvalues. Their eigenvectors are orthogonal with respect to a suitably defined $CPT$ inner product \cite{02bend}. It has recently been conjectured that $PT$-symmetry is closely related to split-quaternionic extensions of quantum theory \cite{11brod, 12sato}. Here we show that split-quaternionic extensions of Hermitian matrices are indeed a natural representation of $PT$-symmetric matrices. This equivalence allows us to introduce new $PT$-symmetric random matrix ensembles.

In conventional quantum systems, random matrices play an important role due to their ability to describe spectral fluctuations in sufficiently complicated systems \cite{Meht69}. In particular, the famous Bohigas-Giannoni-Schmit conjecture states that the spectral fluctuations of quantum systems with chaotic classical counterparts are similar to those of certain random matrices \cite{84bohi,Haak92}. There are three important universality classes for Hermitian quantum systems, depending on the time-reversal properties of the system, corresponding to the Gaussian orthogonal, unitary, and symplectic ensembles \cite{62dyso}. Non-Hermitian random matrix models, on the other hand, whose eigenvalues are in general complex, are widely studied, and have applications ranging from dissipative quantum systems and scattering theory to quantum chromodynamics (see, e.g., \cite{Khor11} and references therein). Several attempts towards defining $PT$-symmetric random matrices and identifying universality classes for $PT$-symmetric systems have been made \cite{Ahme03,10Wang,Gong12,13sriv,13ahme,Bohi13}. Most of them are restricted to $2\times 2$ matrices, due to the lack of a natural parameterisation of larger $PT$-symmetric matrices. Here we introduce the split-complex and split-quaternionic versions of the Gaussian unitary and symplectic ensembles as candidates for representing new universality classes for $PT$-symmetric quantum systems. The new ensembles are invariant under orthogonal and unitary transformations, reflecting the invariance of split-complex and split-quaternionic Hermitian matrices, respectively. We show that the split-complex case is closely related to the well-known real Ginibre ensemble \cite{65gini,Khor11}, which in itself is $PT$-symmetric. The formulation in terms of split-complex Hermitian matrices might offer a new approach towards solving some outstanding problems regarding the spectral features of Ginibre matrices. To the best of our knowledge the split-quaternionic Hermitian ensemble is not directly related to any previously studied ensemble. 

Let us start with a brief summary of some important properties of split-complex and split-quaternionic numbers and matrices that we will need in the following. Split-complex numbers \cite{68yagl} are numbers of the form $z = x + \uj y$, where $x, y \in\mathds{R}$ and $\uj$ is the imaginary unit of the algebra, such that $\uj^2 = 1.$ The conjugate of a split-complex number is given by $\overline{z} = x - \uj y$.
The ``norm'' of a split complex number, defined by $|z|^2=z\bar z=x^2-y^2,$
can be positive, null, or negative. 
A split-complex number can be represented as a real $2\times 2$ matrix
\begin{equation}
\label{split_quat_matrix_rep}
z\leftrightarrow\left(\begin{array}{cc} x& y\\ y&x\end{array}\right).
\end{equation}
Similarly, a generic split-quaternion \cite{49cock} $p\in\HH_S$ can be written as 
$p = p_0 + \ui p_1 +\uj p_2 + \uk p_3$,
where $p_i\in\RR$, and $\ui, \uj, \uk$ satisfy the relations 
\begin{equation}
\ui^2 = -1,\quad \uj^2 = \uk^2 = \ui\uj\uk = +1.
\end{equation}
The conjugate of a split-quaternion is defined as 
$
\overline{p} = p_0 - \ui p_1 - \uj p_2 -\uk p_3$. The ``norm'' of a split-quaternion is defined similar to the split-complex case by 
$
|p|^2 = \bar p p=p_0^2+p_1^2-p_2^2-p_3^2,
$
and again can be positive, null, or negative. A split-quaternion can be represented as a complex $2\times 2$ matrix
\begin{equation}
\label{split_quat_matrix_rep}
p\leftrightarrow\left(\begin{array}{cc} p_0+\ui p_1& p_2+\ui p_3\\ p_2-\ui p_3&p_0-\ui p_1\end{array}\right).
\end{equation}

In analogy to the case of quaternionic vector spaces \cite{14rodml} we can define an (indefinite) inner product for two vectors $\vec{u},\vec{v}\in\mathds{H}_S^N$ with split-quaternionic components as 
\begin{equation}
\label{eqn_splitprod}
(\vec{u},\vec{v})=\sum_{n=1}^N \bar u_n v_n. 
\end{equation}
The adjoint $A^\dagger$ of a split-quaternionic matrix $A$ is then defined as $(\vec{u},A\vec{v})=(A^\dagger\vec{u},\vec{v})$, which in terms of the elements means taking the transpose and split-quaternionic conjugate, 
$A^\dagger = (a^\dagger_{jk}) = (\overline{a_{kj}})$.
We can generalise the concept of Hermiticity to split-quaternionic matrices, by referring to matrices with split-quaternionic elements that satisfy $H^\dagger = H$ as \textit{split-Hermitian}.
The set of split-Hermitian matrices is invariant under standard unitary transformations. The subset of split-complex Hermitian matrices is invariant under orthogonal transformations only. The space of $N\times N$ split-Hermitian matrices is $(2N^2-N)$-dimensional, and the subspace of split-complex Hermitian matrices is $N^2$-dimensional. 

We can define eigenvalues and eigenvectors of split-Hermitian matrices using the complex $2\times 2$ matrix representation in (\ref{split_quat_matrix_rep}) for the elements. The resulting characteristic polynomial is real, thus the eigenvalues are either real or come in complex conjugate pairs. Further, the eigenvalues of the $2N\times2N$ representation are doubly degenerate, in analogy to Kramer's degeneracy for quaternionic Hermitian matrices \cite{Meht69,Haak92,12sato}. The split-quaternionic eigenvectors belonging to distinct eigenvalues are orthogonal with respect to the inner product (\ref{eqn_splitprod}).

Following \cite{10bend} we define the class of $PT$-symmetric matrices as the set of complex matrices with real characteristic polynomial. Thus, any split-Hermitian matrix can be interpreted as a $PT$-symmetric Hamiltonian. A simple counting argument shows that the spaces of split-Hermitian and complex $PT$-symmetric matrices are in fact isomorphic. An $N \times N$ complex matrix has $2N^2$ real parameters. The coefficients of the characteristic polynomial are linearly independent linear functions of the parameters. That is, if we condition the matrix to have a real characteristic polynomial, we find one constraint for the parameters for the coefficient of each power in the polynomial, yielding $N$ linearly independent constraints. That is, a $PT$-symmetric matrix can be parameterised by $2N^2-N$ parameters, just as a split-Hermitian matrix of the same size, thus they are isomorphic. In particular, the subspace of split-complex Hermitian matrices is isomorphic to the space of real $PT$-symmetric matrices (which is just the space of all real matrices). 

Let us now introduce random matrix ensembles of split-complex and split-quaternionic Hermitian matrices, as direct generalisations of the standard Gaussian unitary and Gaussian symplectic ensembles. That is, we consider split-quaternionic and split-complex matrices whose elements are independently distributed normal random variables, subject to the constraint of split-Hermiticity. We define the probability density function on the space of split-complex Hermitian matrices $H$ as 
\begin{equation}
\label{eqn_P(H)_compl}
\mathbb{P}(H)\ud H = \big(\frac{1}{\pi}\big)^{\frac{N}{2}}\big(\frac{2}{\pi}\big)^{\frac{1}{2}N(N-1)}\ue^{-\mathrm{Tr}(HH^T)}\ud H,
\end{equation}
where $ \ud H = \prod_{m < n}\ud \Re(h_{mn}) \ud \Im(h_{mn})\prod_{m=1}^N \ud h_{mm}.$
Thus, this ensemble is invariant under orthogonal transformations of $H$.

In the split-quaternionic case, we define the probability distribution on the space of matrices $H$ as
\begin{equation}\hspace{-0.1pt}
\label{eqn_P(H)_quat}
\uP(H)\ud H = \big(\frac{2}{\pi}\big)^{\frac{N}{2}}\big(\frac{2}{\sqrt{\pi}}\big)^{2N(N-1)}\ue^{-\mathrm{Tr}(HH^{I} + H^{I}H)}\ud H, \hspace{-5pt}
\end{equation}
where  $ \ud H = \prod_{m < n}\hspace{-2.5pt}\ud h^1_{mn} \ud h^{\ui}_{mn}\ud h^{\uj}_{mn} \ud h^{\uk}_{mn}\prod_{m=1}^N\hspace{-2.5pt} \ud h_{mm},$ and here $H^{I}$ performs the transpose of $H$ and complex conjugation with respect to the imaginary unit $\ui$ only. Hence this ensemble is invariant under unitary transformations of $H$, reflecting the invariance class of split-quaternionic Hermitian matrices. Note that the matrix elements of $H$ are by definition statistically independent for both ensembles (\ref{eqn_P(H)_compl}) and (\ref{eqn_P(H)_quat}). That is, these new ensembles fulfil the two defining properties of the standard Gaussian ensembles, of independence of the matrix elements and invariance under the invariance transformations of the class of matrices on which the ensemble is defined. We conjecture that the two split-Hermitian ensembles constitute universality classes for $PT$-symmetric quantum systems.

We shall now derive analytic expressions for the joint probability density of the eigenvalues and the one-level densities for the case of $N=2$, following a method similar to the one used in \cite{08somm} for the real Ginibre ensemble. Let us start with the split-complex case. Writing
\begin{equation}
H = \left(\begin{array}{cc} \Lambda_1 & \delta - \uj \gamma \\ \delta + \uj \gamma & \Lambda_2 \end{array}\right),
\end{equation} 
with $\Lambda_{1,2}, \delta, \gamma \in \mathds{R}, \uj ^2 = 1$, we have
\begin{equation}
\uP(H) =  \frac{2}{\pi^2}\ue^{-\mathrm{Tr}(HH^T)} = \frac{2}{\pi^2}\ue^{-(\Lambda_1^2 + \Lambda_2^2 + 2\delta^2 + 2\gamma^2)}.
\end{equation}
The eigenvalues of H are given by
\begin{equation}
\lambda_{1,2} = \frac{\Lambda_1 + \Lambda_2}{2} \pm \sqrt{\Big{(}\frac{\Lambda_1 - \Lambda_2}{2}\Big{)}^2 + \delta^2 - \gamma^2}.
\end{equation}
Clearly these eigenvalues can be real or complex conjugate, depending on the values of $\delta$ and $\gamma$.

To deduce the joint probability density of the eigenvalues $\lambda_{1,2}$ we have to transform from $\Lambda_{1,2},\, \gamma,$ and $\delta$ to $\lambda_{1,2},\, \gamma,$ and $\delta$, and then integrate over $\gamma$ and $\delta$. We have 
\begin{equation}
\Lambda_{1,2} = \frac{\lambda_1 + \lambda_2}{2} \pm \sqrt{\Big{(}\frac{\lambda_1 - \lambda_2}{2}\Big{)}^2 - \delta^2 + \gamma^2},
\end{equation}
and thus we find  
\begin{equation}
\uP(\lambda_{1,2}, \delta, \gamma) = \frac{1}{\pi^2}\frac{|\lambda_1 - \lambda_2|\ue^{-(\lambda_1^2 + \lambda_2^2 + 4\gamma^2)}}{\sqrt{\Big{(}\frac{\lambda_1 - \lambda_2}{2}\Big{)}^2 - \delta^2 + \gamma^2}}.
\end{equation}
When integrating over $\delta$ and $\gamma$ we have to consider that $\Lambda_1, \Lambda_2,\delta, \gamma$ are real by definition. That is, we integrate over the regions $\delta^2 \leq \Big{(}\frac{\lambda_1 - \lambda_2}{2}\Big{)}^2 + \gamma^2$, and $\gamma^2 \geq - \Big{(}\frac{\lambda_1 - \lambda_2}{2}\Big{)}^2$ to find
\begin{equation}
\label{eqn_jpdf_sc}
\uP(\lambda_1, \lambda_2) = \frac{\ue^{-(\lambda_1^2 + \lambda_2^2)}|\lambda_1 - \lambda_2|\mathrm{erfc}(2|\Im(\lambda_{1,2})|)}{2\sqrt{\pi}}.
\end{equation}

From the joint probability distribution we can deduce the one-level density of the eigenvalues. For this purpose we have to distinguish between the case of real or complex conjugate eigenvalues. If $\lambda_1, \lambda_2$ are real then $\mathrm{erfc}(0)=1.$
Integrating over one of the real eigenvalues then yields the one-level density for real eigenvalues 
\begin{equation}
R_1^{\mathds{R}}(\lambda) = R_1(\lambda | \lambda \in \mathds{R}) = \frac{\lambda \ue^{-\lambda^2}}{2}\mathrm{erf}(\lambda) +  \frac{\ue^{-2\lambda^2}}{2\sqrt{\pi}}. \label{GSUEreal}
\end{equation}
We see that there is a non-zero probability to obtain two real eigenvalues, which is given by the normalisation of $R_1^{\mathbb{R}}(\lambda)$, as $\frac{1}{\sqrt{2}}$.
\\If $\lambda_1, \lambda_2$ are complex conjugate one of the eigenvalues entirely determines the other, and the joint probability density automatically reduces to the one-level density for complex eigenvalues
\begin{equation}
R_1^{\mathds{C}}(\lambda)\! =\! \frac{2|\Im(\lambda)|}{\sqrt{\pi}}\ue^{-2(\Re(\lambda)^2\! -\! \Im(\lambda)^2)}\mathrm{erfc}(2|\Im(\lambda)|). \label{GSUEcompl}
\end{equation}
The factor of two accountes for the fact that the eigenvalues are indistinguishable.
The total one-level density of a complex eigenvalue is then given by
\begin{equation}
R_1(\lambda) = R_1^{\mathds{C}}(\Re(\lambda), \Im(\lambda)) + \delta(\Im(\lambda))R_1^{\mathds{R}}(\Re(\lambda)). \label{totaldens}\end{equation}

\begin{figure}
\includegraphics[width=0.99\textwidth]{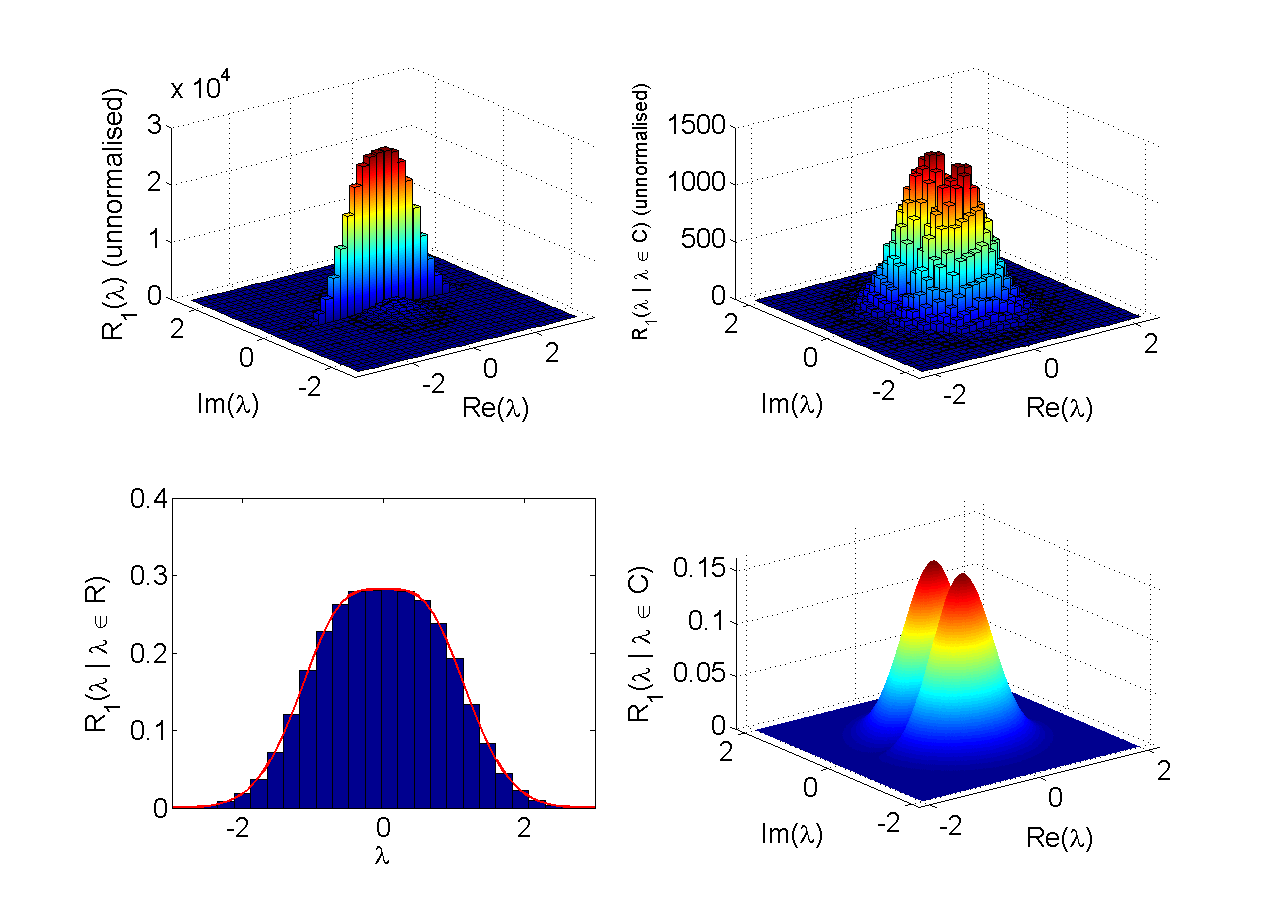}    
\vspace{-15pt}
\caption{One-level density for the $2\times2$ split-complex Hermitian ensemble (\ref{eqn_P(H)_compl}). The top left panel shows the histogram of the eigenvalues using numerically obtained eigenvalues from $200,000$ matrices. The right panels show the histogram of the conditionally complex eigenvalues (top) in comparison to the analytical result (bottom). The bottom left panel shows the numerical and analytical distribution for conditionally real eigenvalues.}
\centering
 \label{splitcompdists} 
\end{figure}

Figure \ref{splitcompdists} depicts the numerically obtained histogram of the eigenvalues of a large number of split-complex Hermitian matrices sampled from the distribution (\ref{eqn_P(H)_compl}). We also plot the probabilities conditional on real and complex eigenvalues respectively in comparison with the analytical results. We observe a good agreement between the numerics and the analytical results; in particular, there is a sharp peak in the histogram for $\Im(\lambda) = 0,$ reflecting the $\delta$-function in the analytic distribution.

Note that the distributions (\ref{GSUEreal}) and (\ref{GSUEcompl}) are similar to those obtained from the $2\times 2$ real Ginibre ensemble \cite{08somm}. In fact, the two ensembles can be explicitly related. The $2 \times 2$ split-complex Hermitian ensemble can be brought, via a parameter-independent orthogonal transformation on the $4 \times 4$ matrix representation, to a matrix which is spectrally equivalent to two copies of a matrix from the $2 \times 2$ real Ginibre ensemble, with elements distributed as $\mathcal{N}(0,\frac{1}{2})$ instead of $\mathcal{N}(0,1)$. Explicitly we find that
\begin{equation} 
O^T\left(\begin{array}{cccc} \Lambda_1 & 0 & \delta & \gamma \\
                          0 & \Lambda_1 & \gamma & \delta  \\
                         \delta & -\gamma & \Lambda_2 & 0 \\
                         -\gamma & \delta & 0 & \Lambda_2 \end{array}\right)O =\left(\begin{array}{cccc} a & b & 0 & 0 \\ d & c & 0 & 0 \\ 0 & 0 & a & d \\ 0 & 0 & b & c \end{array}\right), 
\end{equation}
with
\begin{equation}
O = \frac{1}{\sqrt{2}}\left(\begin{array}{cccc}0 & 1 & 0 & -1 \\ 0 & 1 & 0 & 1 \\ 1 & 0 & -1 & 0 \\ 1 & 0 & 1 & 0 \end{array}\right),\end{equation} 
and $a = \Lambda_2$, $b = \delta - \gamma$, $d = \delta+\gamma$, and $c = \Lambda_1$. That is $a,b,c,d$ are independently and identically distributed as $\mathcal{N}(0, \frac{1}{2})$. A similar relation holds in the $N\times N$ case.

Let us now consider the split-quaternionic case. Let 
\begin{equation}
H = \left(\begin{array}{cc} \Lambda_1 & \delta -\ui\mu-\uj\gamma-\uk\sigma \\ \delta + \ui\mu+\uj\gamma+\uk\sigma & \Lambda_2 \\ \end{array}\right),
\end{equation}
with $\Lambda_{1,2}, \delta, \mu, \gamma, \sigma \in \mathds{R}$; that is, we have the probability distribution
\begin{equation}
\mathbb{P}(\Lambda_{1,2},\delta,\mu,\gamma,\sigma) =\frac{32}{\pi^3}\ue^{-2(\Lambda_1^2 + \Lambda_2^2 + 2(\delta^2+\mu^2+\gamma^2+\sigma^2))}. 
\end{equation}
Transforming to the variables $\lambda_{1,2}$, $\delta$, $\mu$, and $\gamma$, we obtain
\begin{equation}\hspace{-0.1pt}\mathbb{P}(\lambda_{1,2},\delta,\mu,\gamma,\sigma) = \frac{32\ue^{-2(\lambda_1^2 + \lambda_2^2 + 4(\gamma^2 + \sigma^2))}|\lambda_1-\lambda_2|}{\pi^3\sqrt{(\lambda_1-\lambda_2)^2+4\Omega^2}},\hspace{-5pt}\end{equation}
where $\Omega^2 = \gamma^2+\sigma^2-\delta^2-\mu^2.$
We now proceed to integrate over the regions where the original parameters remain real. To ensure that $\Lambda_{1,2} \in \mathds{R}$ we need $\frac{(\lambda_1-\lambda_2)^2}{4} + \Omega^2 \geq 0.$
Rearranging this inequality as a limit on $\mu$ and integrating over $\mu$, followed by integrating over $\delta$ over the region $\delta^2 \leq \frac{(\lambda_1-\lambda_2)^2}{4}+\gamma^2+\sigma^2$, leaves
\begin{equation}
\mathbb{P}(\lambda_{1,2},\gamma,\sigma) = \frac{32|\lambda_1-\lambda_2|}{\pi^2}\ue^{-2(\lambda_1^2 + \lambda_2^2 + 4(\gamma^2 + \sigma^2))}\mathcal{E}(\lambda_{1,2}),
\end{equation}
where $\mathcal{E}(\lambda_{1,2})^2 = \frac{(\lambda_1-\lambda_2)^2}{4}+\gamma^2+\sigma^2.$ To ensure that $\delta \in\mathds{R}$, we additionally require
$\mathcal{E}(\lambda_{1,2})^2 \geq 0.$
Now, transforming our variables $(\gamma,\sigma)$ into polar coordinates $(r,\phi)$ and integrating over
$\phi \in [0, 2\pi]$ leaves the joint probability density function of $\lambda_1, \lambda_2$,
\begin{equation}\mathbb{P}(\lambda_1, \lambda_2) = \frac{64|\lambda_1-\lambda_2|}{\pi}\ue^{-2(\lambda_1^2 + \lambda_2^2)} \mathcal{R}(\lambda_{1,2}), \label{GSSEjointdist} \end{equation}
where $\mathcal{R}(\lambda_{1,2}) = \int_{\Gamma}r \ue^{-8r^2}\sqrt{\frac{(\lambda_1 - \lambda_2)^2}{4} + r^2}\ud r.$ The region $\Gamma$ is different for real eigenvalues in comparison to complex conjugate eigenvalues.

In the case of real eigenvalues, the reality of $\delta$ implies that we integrate over $r \in [0, \infty].$ We thus find the joint probability distribution for real eigenvalues as
\begin{eqnarray}
\mathbb{P}^{\mathds{R}}(\lambda_1,\lambda_2) &= \frac{2}{\pi}(\lambda_1-\lambda_2)^2\ue^{-2(\lambda_1^2+\lambda_2^2)} \nonumber \\
&+ \frac{|\lambda_1-\lambda_2|}{\sqrt{2\pi}}\ue^{-4\lambda_1\lambda_2}\mathrm{erfc}(\sqrt{2}|\lambda_1-\lambda_2|) \label{GSSErealjoint}.
\end{eqnarray}
Integrating over one of the eigenvalues we find the one-level density in the case of real eigenvalues as 
\begin{equation}
R_1^{\mathds{R}}(\lambda)=\frac{\ue^{-4\lambda^2}}{8\lambda^2\sqrt{2\pi}} + \frac{\ue^{-2\lambda^2}}{\sqrt{2\pi}}\Big(2\lambda^2+1 - \frac{1}{8\lambda^2}\Big) . \label{GSSEreal} \end{equation}
In the split-quaternionic case, the probability of obtaining real eigenvalues is $1 - \frac{1}{2\sqrt{2}}$, slightly smaller than in the split-complex case. 

If $\lambda_1$ and $\lambda_2$ are complex conjugate, however, the reality of $\delta$ means that $\gamma^2 + \sigma^2 \geq \Im(\lambda_{1,2})^2.$
That is, we now have to integrate over $r \in [|\Im(\lambda_{1,2})|, \infty]$. 
Accounting for the fact that the eigenvalues are indistinguishable, we thus obtain the one-level density for complex eigenvalues
\begin{equation}R_1^{\mathds{C}}(\lambda) = 2\sqrt{\frac{2}{\pi}}|\Im(\lambda)|\ue^{-4\big((\Re(\lambda))^2 + (\Im(\lambda))^2\big)}.\end{equation}
Again we can express the total one-level density of a complex eigenvalue in the form (\ref{totaldens}).

\begin{figure}
\includegraphics[width=0.99\textwidth]{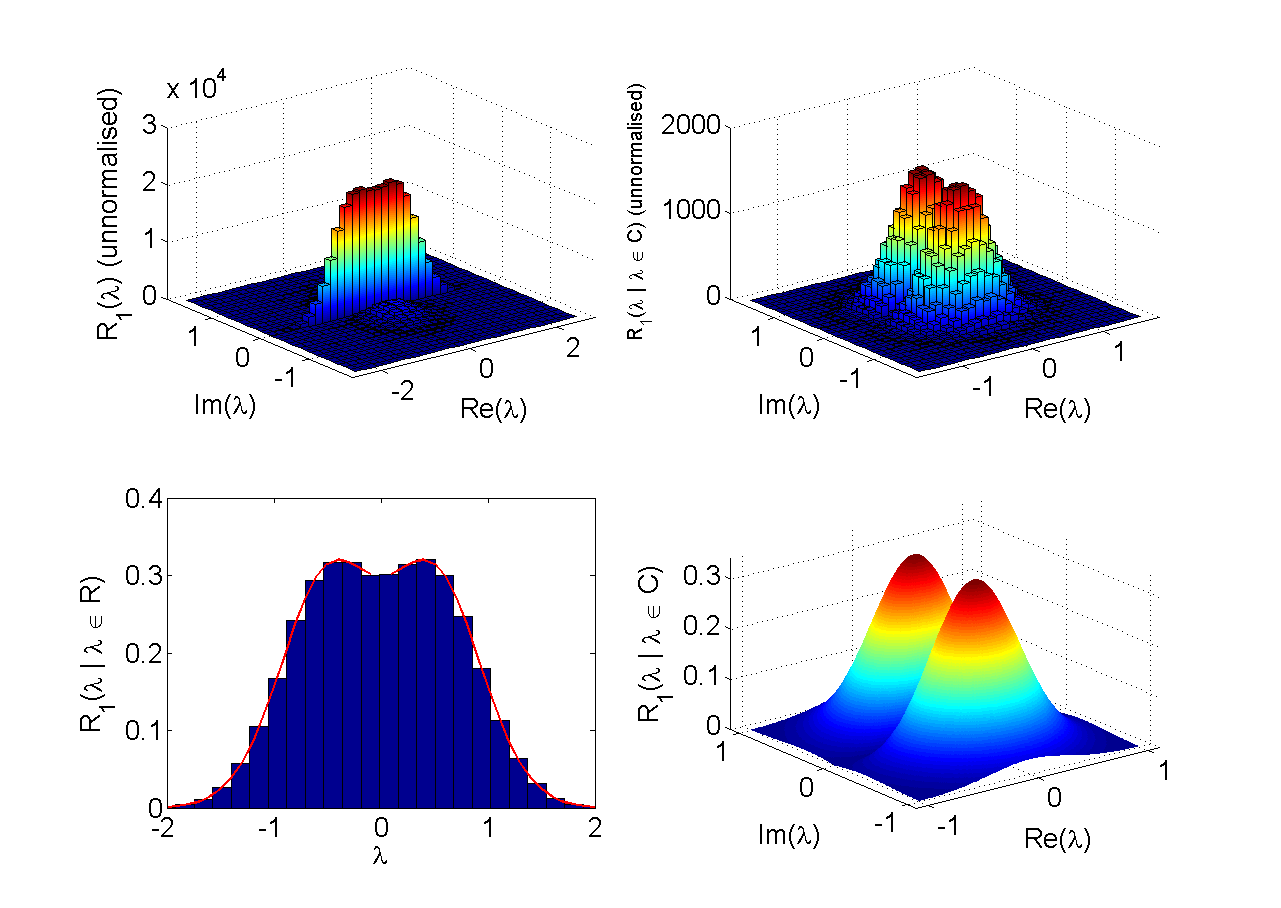}  
\vspace{-15pt}
\caption{As figure \ref{splitcompdists} for the split-quaternionic case.}
\centering
 \label{fig_eig_hist_quat} 
\end{figure}

Figure \ref{fig_eig_hist_quat} depicts numerically obtained eigenvalue densities for a large number of matrices sampled from the distribution (\ref{eqn_P(H)_quat}) in comparison with the analytical results. We observe a dip in the one-level density for real eigenvalues, similar to the one in the $GUE$ one level density for $2\times2$ matrices. The exact form of the distribution (\ref{GSSEreal}) however, is distinctively different from the GUE case.

Let us finally derive the level spacing distributions in the case of real eigenvalues for the split-Hermitian ensembles.
These are the distributions of the spacings $s \propto |\lambda_1 - \lambda_2|$, subject to the requirement that the mean level spacing is one. 
For this purpose we subsitute $\mu_1 = \lambda_1 - \lambda_2, \hspace{3pt}\mu_2 = \lambda_1 + \lambda_2$, in the joint probabilities of the real eigenvalues. For the split-complex case this yields
\begin{equation}\uP(\mu_{1,2}) = \frac{1}{4\sqrt{\pi}}|\mu_1|\ue^{-\frac{1}{2}(\mu_1^2 + \mu_2^2)}.\end{equation}
Integrating $\mu_2$ over the whole real line, normalising and rescaling to ensure that the mean spacing is one, we obtain 
\begin{equation} \uP(s) = \frac{\pi}{2}s\ue^{-\frac{\pi}{4}s^2}. \end{equation}
This is identical to the level spacing for the GOE, as has been noted before for the real Ginibre ensemble \cite{Khor11}.

For the split-quaternionic Hermitian ensemble we again substitute $\mu_1$ and $\mu_2$ in the joint probability distribution for real eigenvalues to obtain
\begin{equation}\hspace{-0.1pt}\uP(\mu_{1,2}) = \frac{\mu_1^2}{\pi}\ue^{-\mu_1^2 - \mu_2^2} + \frac{|\mu_1|}{2\sqrt{2\pi}}\ue^{-\mu_2^2 + \mu_1^2}\mathrm{erfc}(\sqrt{2}|\mu_1|). \hspace{-5pt}\end{equation}
After integrating over $\mu_2$ and appropriate rescaling we obtain
\begin{equation} \uP(s)= 
\frac{4\sqrt{2a^3}\Big(\frac{s^2}{\sqrt{\pi}}\ue^{-as^2} + \frac{1}{2\sqrt{2}}s\ue^{as^2}\mathrm{erfc}(\sqrt{2a}s)\Big)}{2\sqrt{2} - 1}, \end{equation}
where
$a = \Big(\frac{3\sqrt{2} - \mathrm{sinh}^{-1}(1)}{(2\sqrt{2} - 1)\sqrt{\pi}}\Big)^2.$

\begin{figure}
\includegraphics[width=0.99\textwidth]{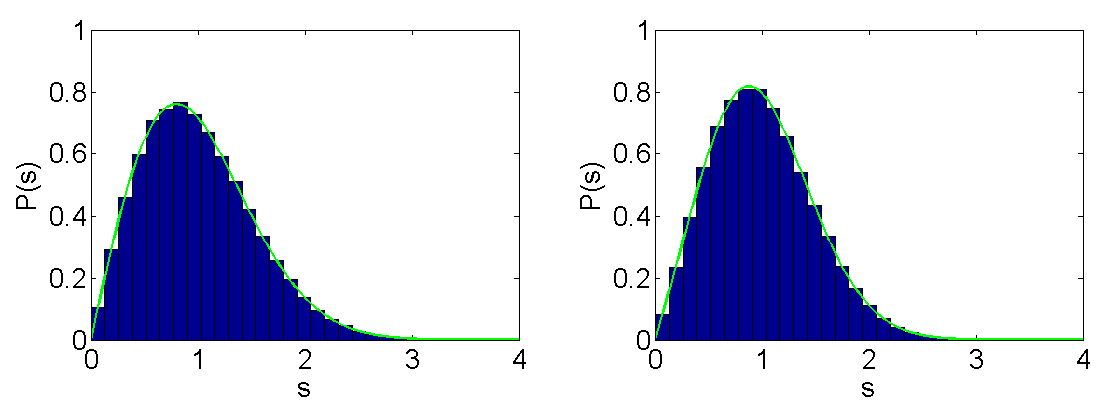}  
\vspace{-10pt}
\caption{Level spacing distributions for real eigenvalues from 100,000 matrices from the split-complex (left) and split-quaternionic (right) ensembles.}
\centering
 \label{GSSElevelreal} 
\end{figure}

Figure  (\ref{GSSElevelreal})  depicts the numerically obtained histogram of the level spacings of the eigenvalues of a large number of split-complex Hermitian and split-quaternionic Hermitian matrices sampled from distributions  (\ref{eqn_P(H)_compl}) and (\ref{eqn_P(H)_quat}). We observe an excellent agreement with the analytical reasults. Note that the level repulsion is linear for both ensembles.

In summary, we have shown that the space of $PT$-symmetric matrices is isomorphic to the space of split-Hermitian matrices. We have introduced two new random matrix ensembles of split-complex and split-quaternionic Hermitian Gaussian random matrices, and derived analytic expressions for the $2\times 2$ case. We have demonstrated a relation between the real Ginibre ensemble and the split-complex Hermitian ensemble. It is a challenging but achievable task to derive analytic results for general $N\times N$ matrices for the new ensembles. The split-complex Hermitian ensemble might help in understanding some of the still-unknown spectral properties of the $N\times N$ real Ginibre ensemble. We have conjectured that the newly derived ensembles constitute universality classes for $PT$-symmetric systems. Hitherto only few $PT$-symmetric model systems have been investigated with respect to their statistical spectral properties. In \cite{09degu}, GOE-type statistics were observed for a $PT$-symmetric Dicke model in the special case of real eigenvalues, which is consistent with the spectral properties of the split-complex ensemble introduced here. It is an important task to compare the spectral features of the new ensembles to those of further $PT$-symmetric model systems. 

\section*{Acknowledgements}
EMG acknowledges support via the Imperial College JRF scheme, the L'Or\'eal UNESCO Women in Science programme, and from the Royal Society. SMN acknowledges support from an EPSRC DTA grant.

\end{document}